\documentclass{vgtc}       

\ifpdf
 \pdfoutput=1\relax     
 \usepackage{graphicx}    
 \DeclareGraphicsExtensions{.pdf,.png,.jpg,.jpeg} 
\else
 \usepackage{graphicx}    
 \DeclareGraphicsExtensions{.eps}  
\fi%

\graphicspath{{figures/}{pictures/}{images/}{./}} 
\usepackage{microtype}     
\PassOptionsToPackage{warn}{textcomp} 
\usepackage{xcolor}
\usepackage{textcomp}     
\usepackage{mathptmx}     
\usepackage{times}      
\usepackage{cite}      
\usepackage{tabu}      
\usepackage{booktabs}     
\usepackage[bottom]{footmisc}
\usepackage{caption}
\usepackage[utf8]{inputenc}
\usepackage{array}
\usepackage{wrapfig}
\usepackage{multirow}
\usepackage{tabu}

\usepackage{graphicx}

\onlineid{0}
\vgtccategory{Research}
\vgtcinsertpkg

\title{Show or Tell? Visual and Verbal Representations Bias Position Recall}

\author{Cristina R. Ceja \thanks{Cristina R. Ceja is with Northwestern University. Email: crceja@u.northwestern.edu}
\and Cindy Xiong\thanks{Cindy Xiong is with University of Massachusetts Amherst. Email: cindy.xiong@cs.umass.edu} %
}

\abstract{
When we view visualizations, we not only have a visual representation of the data, but also a verbal one. Recent work has shown that these visual representations of data can be biased, such that the position of a line in a chart will be consistently underestimated. But are the verbal representations of position encodings also biased in the same manner, or is this a purely visual bias that can be mitigated with verbal context? We explored the bias in position reproductions for simple uniform lines for both visual and verbal representations. We find that the direction of the bias changed depending on the response modality, with visual reproductions showing a position underestimation while verbal responses showed overestimation. This finding indicates that, even for simple line charts, biases are still present for both visual and verbal representations, although the directionality of this bias depends on the modality.
}

\keywords{
Biases, line charts, response modality, position estimation.
}

\begin{document}
\maketitle
\section{Introduction}
How good are we at remembering the values we see in a visualization? Recent work has shown that the visual recall of data in even simple line charts can be systematically biased. When viewers were asked to recreate the vertical position of a line mark with a uniform distribution by visually dragging a line probe to the recalled position, viewers consistently recalled the position of the line to be \textit{lower} than it actually appeared \cite{xiong:2020:PerceptualPull}. This consistent underestimation was claimed to indicate that our visual memory for position encodings in line charts is systematically biased, such that we will consistently underestimate the position of a line mark.

However, in the real world, visualization viewers will seldom need to visually recreate a chart to later interpret it. Imagine viewing a line chart depicting the sales trends for a given quarter. You might want to eyeball the average sales to later compare it to the number from the previous quarter. In this process, your recall of the average numerical value is not a visual representation, but rather a verbal one (e.g., 42). 


Since previous work only tested visual representations with a visual reporting modality, it is unknown whether the verbal representations of a visual encoding within a visualization are also biased. Visual images, including charts and diagrams, can be dual-coded, such that these images are both visually and verbally encoded \cite{paivio:2013:Imagery}. For example, when we view a data visualization, we not only have a visual representation (e.g., the estimated position of the data in space), but also a verbal representation which may convey semantic information about it (e.g., the estimated numerical values of the data when provided the context of y-axis values). Previous work has also shown that verbal labels for an image can bias its memory representation, indicating that both verbal and visual cues can be encoded in our representations of visual images \cite{lupyan:2008:Chair}. Therefore, it could be possible that this bias only exists on a visual level, such that the additional context provided with a verbal representation could mitigate this bias. However, it could also be possible that our representation of data in a visualization is biased regardless of whether it was verbally and/or visually encoded, as our memory is often an imperfect process. 
\newline
\textbf{Contribution:} This work replicates recent findings of visual biases in line charts \cite{xiong:2020:PerceptualPull} and extends the investigation to examine whether this bias is purely visual in nature, or exists in both visual and verbal estimations. Differentiating the effects of verbal and visual representations through the use of only verbal or only visual response modalities could be imperative to not only better understand how different representations can influence the accuracy of data recall for visualizations, but to also inspire future research for designing visualization systems or tools to help mitigate this bias.

\section{Experiment}
We empirically test whether verbal representations of position encodings in line charts are similarly biased to visual representations. To do so, we instructed participants to either use a visual or verbal response modality to recall their visual or verbal representations, respectively.

\subsection{General Stimuli}
All experimental stimuli were created with MATLAB using the Psychophysics Toolbox \cite{brainard:1997:Psychophysics, pelli:1997:Videotoolbox} on an Apple Mac Mini running OS 10.10.5. The monitor was 21-inch with a 1280 x 800 pixel resolution and a 60 Hz refresh rate. The approximate viewing distance was an average of 47 cm. 

\begin{figure}[b!]
 \centering
  \vspace{-2mm}
 \includegraphics[width=\columnwidth]{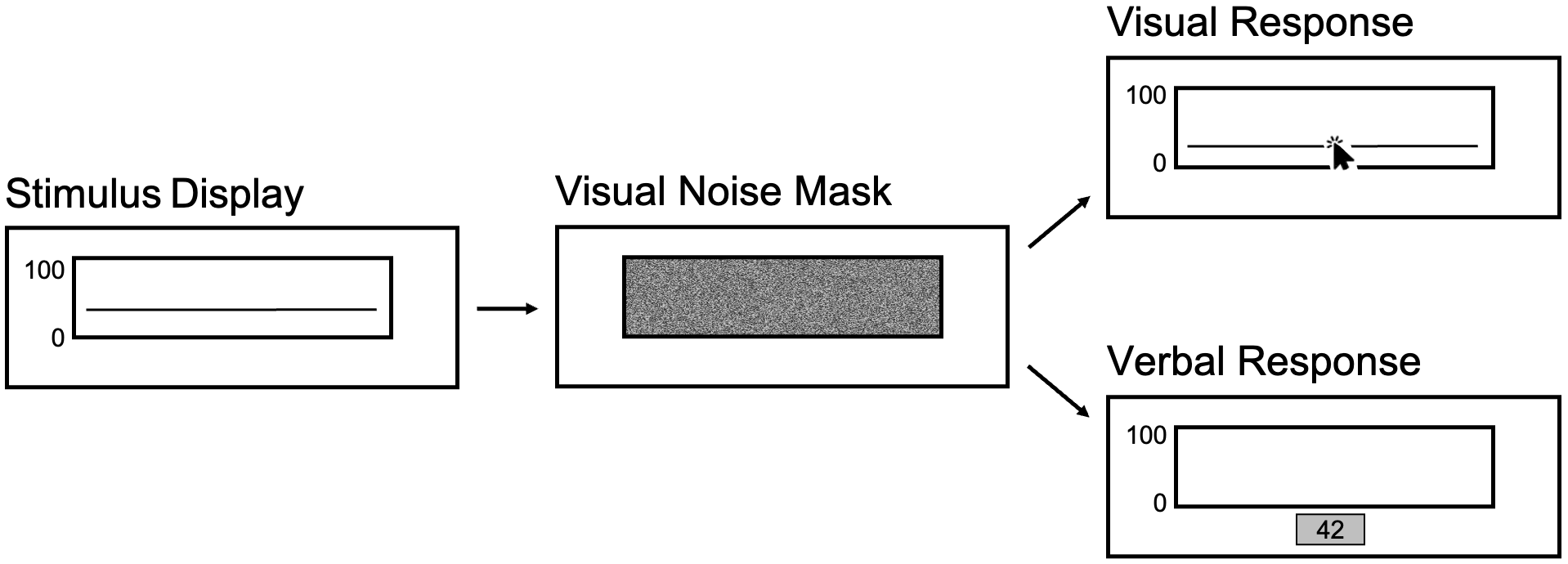}
 \captionsetup{justification = raggedright, singlelinecheck=off}
 \caption{Experimental procedure and design.}
 \label{fig:methods}
\end{figure}

As shown in Figure \ref{fig:methods}, each stimulus display contained a uniform, horizontal line enclosed in a 538 x 140 pixel frame, labeled with the bottom of the frame as 0 and the top of the frame as 100. The vertical position of the line followed a uniform distribution, randomly appearing in any vertical position within the frame.

\begin{figure*}[t!]
 \centering
 \includegraphics[width=18.1cm]{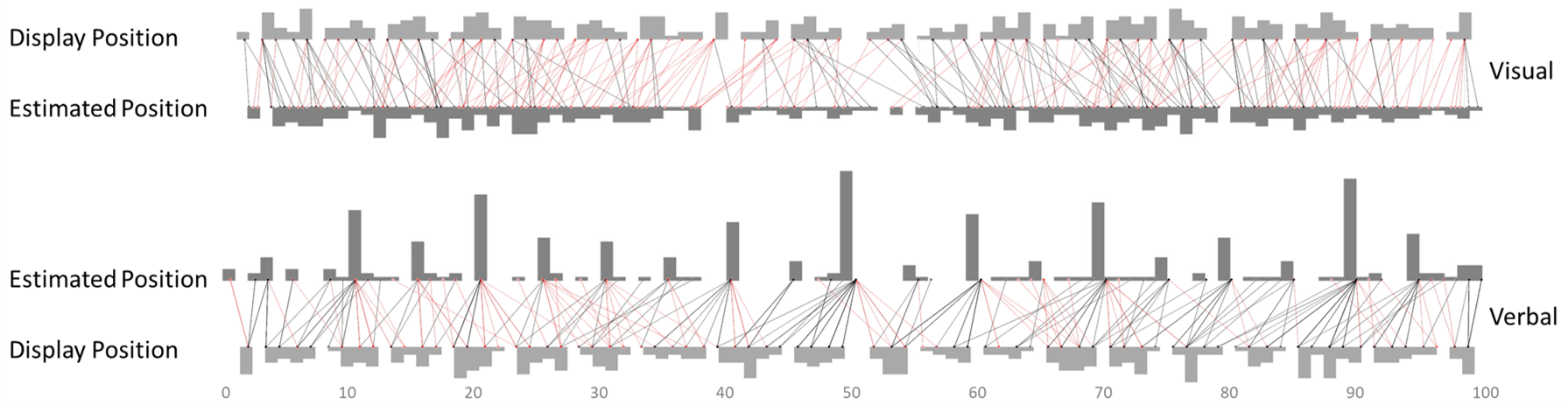}
 
 \caption{Experimental results. The light gray histograms represent the distributions for the displayed vertical positions of the lines, and the dark gray histograms represent the distributions for their estimated vertical position responses. Each line connects a trial's displayed line position to the participant's estimated position for that line, with red lines indicating underestimated responses and gray lines indicating overestimated responses. The top section represents the data for the \textit{Visual} condition where participants responded by dragging a line, and the bottom section represents the data for the \textit{Verbal} condition where participants responded by typing a number between 0 and 100.}
   \vspace{-1mm}
 \label{fig:results}
\end{figure*}

\subsection{Design and Procedure}
Participants saw a stimulus display containing a flat line in a random vertical position on a chart with a y-axis labeled 0 to 100 (see ``Stimulus Display" in Figure \ref{fig:methods}). This was followed by a moving noise mask to reduce any visual afterimages (see ``Visual Noise Mask" in Figure \ref{fig:methods}). Participants were then instructed to estimate the vertical position of the line. In half of all trials, participants made the estimate by dragging a response probe in the form of a horizontal line to report the perceived vertical position of the uniform line (\textit{Visual} condition; see ``Visual Response" in Figure \ref{fig:methods}). In the remaining half of all trials, participants typed a number between 0 and 100 to represent the vertical position of the line (\textit{Verbal} condition; see ``Verbal Response" in Figure \ref{fig:methods}). Conditions were counterbalanced in blocks -- half of the participants saw the \textit{Visual} condition block first, while the remaining half saw the \textit{Verbal} condition block first. Counterbalancing helped to ensure that any observed differences between conditions were not due to learning or fatigue.

Thirteen participants ($M_{age}$ = 18.69, $SD_{age}$ = 0.48, 6 women) from Northwestern University were recruited and completed the experiment for course credit. Each participant completed 24 trials for each condition (\textit{Visual} or \textit{Verbal}), for a total of 48 trials.

\subsection{Results and Discussion}
We excluded the responses with an error (calculated as the difference between response position and actual position) that was greater than 1.5 standard deviation from the mean error for all participants and all trials. With this exclusion criteria, 2.4\% of all responses (or 15 total trials) were excluded from further analysis. As the physical display spanned 140 pixels vertically but participants were shown y-axis labels from 0 to 100 for ease of verbal reporting, we present the following results in both pixel values and in y-axis display values (out of 100) for completeness. 

We observed an overall underestimation of line position when participants estimated the vertical positions by dragging a horizontal line in the \textit{Visual} condition ($M_{error}$ = -0.88 (-1.23 pixels), $SE_{error}$ = 0.28 (0.39 pixels)), replicating findings from \cite{xiong:2020:PerceptualPull}. Interestingly, when participants estimated the vertical position by typing in a number from 0 to 100 in the \textit{Verbal} condition, they overestimated the vertical positions of the lines ($M_{error}$ = 1.82 (2.55 pixels), $SE_{error}$ = 0.32 (0.45 pixels)). A mixed-effects linear model comparing the amount of estimation error between the \textit{Visual} and \textit{Verbal} conditions reveals that the difference between estimations made with the two response modalities is significantly different ($Est$ = 1.92, $p <$ 0.001). 

Additionally, in contrast with the more uniform distribution of the visual reports for line position estimates in the \textit{Visual} condition, the verbal reports of line position estimates in the \textit{Verbal} condition tended to be  rounded-numbers, such as `10's' (57\% of responses; e.g., 50), followed by numbers that ends with `5' (22\% of responses; e.g., 15) (see Figure \ref{fig:results}). To investigate whether this tendency to round estimates to the nearest `5's' or `10's' drove the overestimation for the verbal response modality, we compared estimation error grouped by the last digit of the actual display location value, which ranged from 0 to 9. The overestimation in the verbal condition appeared consistent across all digits, so the overestimation bias was likely not due to a rounding bias, but rather a result of the verbal response modality itself.


\section{Conclusion}
Previous work has shown that visual representations of the position of data in a line chart can be biased, such that it will be consistently underestimated \cite{xiong:2020:PerceptualPull}. However, it was not known whether such biases extended to other representation types, such as verbal representations, or if this bias was unique only to visual ones. The present experiment shows that different response modalities (in this case, visual and verbal response modalities that capture visual and verbal representations, respectively) do result in differing biases for the recall of data values in a line chart. Replicating previous findings, \textit{visual} recall of data in a line chart resulted in a consistent \textit{underestimation} bias. \textit{Verbal} recall of the same data distributions, on the other hand, resulted in an \textit{overestimation} bias. These biases present an important consideration for both visualization design and research. Visualization designers could explicitly annotate the key values in their visualization to help mitigate these biases. Meanwhile, visualization researchers should think carefully about how they want to elicit responses from participants, as the response modality might influence the representation created by the viewer and, consequently, impact the outcome of the study.

Future work can also explore the reasons why different types of representations bias position estimations in opposite directions. What is it about encoding position visually that results in viewers consistently underestimating its value, while encoding position verbally results in consistent overestimation? Would these biases still be present when viewers must both visually and verbally encode the data, or would the respective under- and overestimation cancel out? It would also be important to explore whether these biases persist for other tasks that do not involve recall of data, such as identifying patterns from visualizations or comparing values between two charts.

\bibliographystyle{abbrv-doi}
\bibliography{template}

\begin{thebibliography}{1}

\bibitem{brainard:1997:Psychophysics}
D.~H. Brainard.
\newblock The psychophysics toolbox.
\newblock {\em Spatial Vision}, 10:433--436, 1997.

\bibitem{lupyan:2008:Chair}
G.~Lupyan.
\newblock {From Chair to ``Chair": A Representational Shift Account of Object
  Labeling Effects on Memory}.
\newblock {\em Journal of Experimental Psychology: General}, 137(2):348, 2008.

\bibitem{paivio:2013:Imagery}
A.~Paivio.
\newblock {\em {Imagery and Verbal Processes}}.
\newblock Psychology Press, 2013.

\bibitem{pelli:1997:Videotoolbox}
D.~G. Pelli.
\newblock The videotoolbox software for visual psychophysics: Transforming
  numbers into movies.
\newblock {\em Spatial Vision}, 10:437--442, 1997.

\bibitem{xiong:2020:PerceptualPull}
C.~Xiong, C.~R. Ceja, C.~Ludwig, and S.~Franconeri.
\newblock {Biased Average Position Estimates in Line and Bar Graphs:
  Underestimation, Overestimation, and Perceptual Pull}.
\newblock {\em IEEE Transactions on Visualization and Computer Graphics,
  InfoVis}, 2020.

\end{thebibliography}

\end{document}